\documentclass[twocolumn,printnumbers,amsmath,amssymb,showpacs]{revtex4}

\newcommand{\be}{\begin{equation}}
\newcommand{\ee}{\end{equation}}
\newcommand{\ba}{\begin{eqnarray}}
\newcommand{\ea}{\end{eqnarray}}
\usepackage{graphicx}
\usepackage{dcolumn}
\usepackage{bm}

\begin{document}

\title{Effects of particle-size ratio on jamming of binary mixtures}

\author{Ning Xu$^{1,2}$, Emily S. C. Ching$^3$}

\affiliation{$^1$ Department of Physics, The Chinese University of Hong Kong, Shatin, Hong Kong, China\\
    $^2$ Department of Physics, University of Science and Technology of China, Hefei 230026, China\\
    $^3$ Department of Physics and the Institute of Physics, The Chinese University of Hong Kong, Shatin, Hong Kong, China}

\begin{abstract}
We perform a systematic numerical study of the effects of the particle-size ratio $R \ge 1$ on the
properties of jammed binary mixtures.  We find that changing $R$ does not qualitatively affect the
critical scaling of the pressure and coordination number with the compression near the jamming
transition, but the critical volume fraction at the jamming transition varies with $R$. Moreover,
the static structure factor (density correlation) $S(k)$ strongly depends on $R$ and shows
distinct long wave-length behaviors between large and small particles. Thus the previously reported behavior of $S(k)\sim k$ in the long wave-length limit is only a special case in the $R\rightarrow 1$ limit, and cannot be simply generalized to jammed systems with $R>1$.  
\end{abstract}

\pacs{61.43.-j,61.43.Bn,61.43.Fs,81.05.Rm}

\maketitle

\section{Introduction}
\label{intro}

At zero temperature and shear stress, a packing of frictionless spheres undergoes the jamming transition when its volume fraction $\phi$ is varied across a critical value $\phi_c$ \cite{ohern}.  This transition is denoted as Point J in the jamming phase diagram \cite{liu,zhang}.  Although $\phi_c$ may vary with protocols \cite{torquato,chaud}, it has been
demonstrated that the maximally random jammed packings of hard spheres with single size (mono-disperse) usually exist at a well-defined volume fraction $\phi_c\approx 0.64$ \cite{torquato}, which is also the most probable volume fraction for hard spheres to jam at \cite{xu1}.  When the spheres are deformable and interact via pairwise short-range repulsion, a marginally jammed solid is formed when the volume fraction is slightly greater than $\phi_c$.  Marginally jammed solids have multiple critical scalings with $\phi-\phi_c$ of some mechanical and geometrical quantities such as the pressure, bulk modulus, shear modulus, and coordination number (average number of interacting neighbors per particle) \cite{ohern}, implying the criticality of Point J.  Point J is isostatic \cite{ohern}, i.e. the coordination number $z=z_c=2d$, where $d$ is the dimension of space.  The isostaticity is the minimum requirement of particle constraints to maintain mechanical stability and plays a key role in the control of properties of typical jammed systems such as glasses \cite{xu2,xu3,wyart}.  Therefore, understanding Point J is a good starting point to study some long-standing problems, e.g. glass transition \cite{deben} and the unusual thermal properties of glasses \cite{pohl}.

The jamming transition at Point J is puzzling compared to conventional critical phenomena.  For instance, during the jamming transition, there is no apparent structural change.  Consequently, a diverging static correlation length, e.g. the density correlation length present in the long wave-length limit, that can be observed in typical phase transitions is lacking.  However, jammed packings of mono-disperse spheres still exhibit nonanalytic density fluctuations different from simple liquids \cite{donev,silbert}: the static structure factor $S(k)\sim k$ in the long wave-length limit, where $k$ is the magnitude of the wave vector, while in simple liquids $S(k)$ is quadratic with $S(0)>0$.  The vanishing of $S(0)$ indicates that jammed mono-disperse systems are hyperuniform \cite{donev}.

To successfully avoid crystallization, a binary mixture of particles with two sizes (bi-disperse) is usually employed.  Although some mechanical and vibrational properties of marginally jammed solids do not qualitatively depend on the ratio of the two particle sizes, we are short of systematic knowledge about how the particle-size ratio might affect jamming.  For bi-disperse systems, it is also tempted to assume that either the large or the small particles can be taken as representative to describe the properties of the whole system.  For instance, one may assume that $S(k)$ measured for large particles and small particles are identical, so that the observation of $S(k)\sim k$ in jammed mono-disperse systems is universal and all jammed systems are thus hyperuniform.

In this paper, we study whether and how the particle-size ratio affects the properties of jammed binary mixtures.  By increasing the ratio $R$ of the diameter of the two species from $R=1$, we observe an increase of the critical volume fraction $\phi_c$ and a decrease of the bulk modulus accompanied with the increase of rattlers (particles with no interacting neighbors).  Most importantly, the inclusion of bi-dispersity qualitatively changes the static structure factor $S(k)$ especially in the long wave-length limit.  The linear behavior of $S(k)$ in the long wave-length limit is thus only valid for jammed mono-disperse systems.

\section{Method}
\label{method}

The systems studied consist of $N=10,000$ frictionless spheres with the same mass $m$, contained in three dimensional cubic boxes with length $L$ and periodic boundary conditions. Half of the particles have a diameter $\sigma$, while the other half have a diameter $R\sigma$ with $R\ge 1$.  Particles interact via short range harmonic repulsion: the
inter-particle potential $V(r_{ij})=\epsilon\left( 1-r_{ij}/\sigma_{ij}\right)^2/2$ when the separation between particles $i$ and $j$, $r_{ij}$ is less than the sum of their radii $\sigma_{ij}= (\sigma_i + \sigma_j)/2$, and zero otherwise.  This numerical model has been widely used in the studies of jamming, and is a typical model of granular materials, glasses, and foams.  We vary the diameter ratio $R$ from $1$ to $4$ to carry out a systematic study
of its effects on the properties of the jammed systems. We note that most of the reported studies have focused on two specific values of $R$: $1.0$ and $1.4$.  To generate jammed packings of frictionless spheres, we quickly quench random configurations at high temperatures and fixed volume fraction to the local energy minima using L-BFGS minimization routine \cite{lbfgs}.  We tune the volume fraction until a packing with desired pressure is obtained.  In order to reduce quench rate dependence, we start at a volume fraction which is pretty close to the destination, so that tuning the volume fraction does not lead to significant particle rearrangements.  We use $\sigma$, $m$, and $\epsilon$ as the unit of length, mass, and energy.

For each $R$ and desired pressure, we generate $100$ jammed configurations from independent initial conditions.  The pressure is measured from $P=-\frac{1}{3L^3}\sum_{ij}r_{ij}\frac{{\rm d}V_{ij}}{{\rm d}r_{ij}}$, where the sum is
over all interacting particle pairs.  We obtain the average value of the volume fraction $\phi$ and coordination number $z$ over these $100$ configurations at fixed pressure.  The static structure factor is calculated from $S(k)=\frac{1}{N}\left|\sum_j {\rm exp}\left( {\rm i}\vec{k}\cdot \vec{r}_j\right)\right|^2$ and averaged over
configurations, where the sum is over all particles, and $\vec{r}_j$ is the Cartesian coordinate of particle $j$.  Due to the periodic boundary conditions, the wave vector must be chosen as $\vec{k}=\frac{2\pi}{L}\left(n_x \hat{x} + n_y \hat{y} + n_z \hat{z}\right)$ with the magnitude $k=\frac{2\pi}{L}\sqrt{n_x^2+n_y^2+n_z^2}$, where
$n_x,n_y,n_z=0,\pm1,\pm2,...$.  Because of the particle bi-dispersity, we also separately measure the static structure factor for large-large particles $S_{LL}(k)=\frac{2}{N}\left|\sum_l {\rm exp}\left( {\rm i}{\vec k}\cdot {\vec r}_l\right)\right|^2$, small-small particles $S_{SS}(k)=\frac{2}{N}\left|\sum_s {\rm exp}\left( {\rm i}{\vec k}\cdot {\vec r}_s\right)\right|^2$, and large-small particles $S_{LS}(k)=\frac{4}{N}{\rm Re}\left[ \sum_l {\rm exp} \left( {\rm i} {\vec k}\cdot {\vec r}_l\right) \sum_s {\rm exp} \left( -{\rm i} {\vec k}\cdot {\vec r}_s\right)\right]$, where $\sum_l$ and $\sum_s$ are over all large and all small particles, respectively.  It is obvious that $S(k)=\left[S_{LL}(k) + S_{SS}(k) + S_{LS}(k) \right]/2$.

\begin{figure}
\centering
\includegraphics[width=0.45\textwidth]{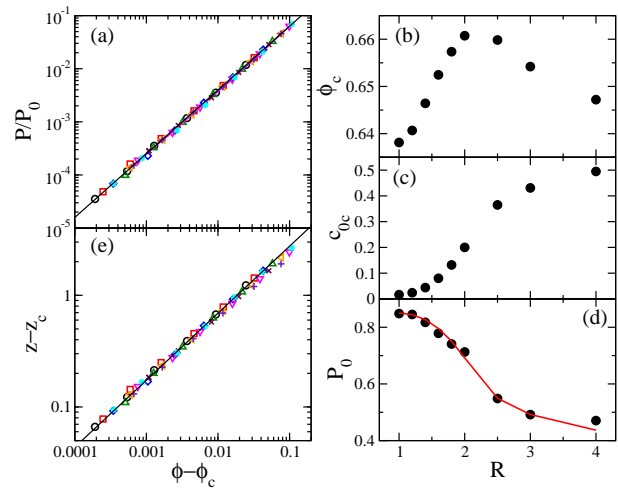}
\caption{\label{fig:fig1}
(a) Rescaled pressure $P/P_0$ and (e) coordination number beyond isostaticity $z-z_c$ versus compression above Point J $\phi-\phi_c$ for marginally jammed solids with the particle size ratio $R=1.0$ (black circles), $1.2$ (red squares), $1.4$ (blue diamonds), $1.6$ (green up triangles), $1.8$ (orange left triangles), $2.0$ (magenta down triangles), $2.5$ (purple pluses), $3.0$ (maroon crosses), and $4.0$ (cyan stars).  The black solid lines in (a) and (e) show $P/P_0=(\phi-\phi_c)^{1.2}$ and $z-z_c=11(\phi-\phi_c)^{0.6}$, respectively.  (b), (c), and (d) show the critical volume fraction at Point J $\phi_c$, concentration of rattlers at Point J $c_{0c}$, and $P_0$ as a function of $R$, respectively.  The red solid line in (d) show $P_0\propto 1-c_{0c}$.  $P_0$ is in the units of $\epsilon/\bar{\sigma}^3$, where $\bar{\sigma}^3=(1+R^3)\sigma^3/2$ is the average volume of particles.
}
\end{figure}

\section{Results and discussion}
\label{results}

Figure~\ref{fig:fig1}(a) shows how the pressure $P$ of marginally jammed solids with different $R$ varies with the volume fraction $\phi$ on the approach of unjamming transition.  The critical volume fraction $\phi_c$ is a fitting parameter to satisfy $P=P_0\left( \phi -\phi_c\right)^a$.  For systems with harmonic repulsion, it is widely accepted that $a=1$ \cite{ohern}.  However, we have noticed in previous studies that $a$ deviates from 1 when $\phi-\phi_c$ is not tiny.  The pressure grows nonlinearly at high compressions, which leads to $a\approx 1.2$ in the range of volume fractions studied, as denoted by the solid line in Fig.~\ref{fig:fig1}(a).

The value of $R$ does affect the jamming transition and material properties of marginally
jammed solids.  When $R$ increases from 1, $\phi_c$ increases and reaches a maximum at
$R\in(2.0,2.5)$, as shown in Fig.~\ref{fig:fig1}(b).  When $R\rightarrow \infty$, the
total volume of small particles is zero, so the system will effectively be mono-disperse
and consequently $\phi_c(\infty) \approx \phi_c(1)$.  It is thus unsurprising that
$\phi_c(R)$ is peaked at an intermediate $R$.  By varying $R$, we obtain about $3.5\%$
increase of $\phi_c$ from mono-disperse systems.  Since Point J controls properties of
jammed solids \cite{xu4} and probably dynamics of supercooled liquids approaching the
glass transition \cite{xu5}, it is worthwhile to investigate further how $R$ affects the dynamics of supercooled colloidal suspensions, e.g. glass transition and equation of state of hard sphere colloids.

\begin{figure}
\centering
\includegraphics[width=0.35\textwidth]{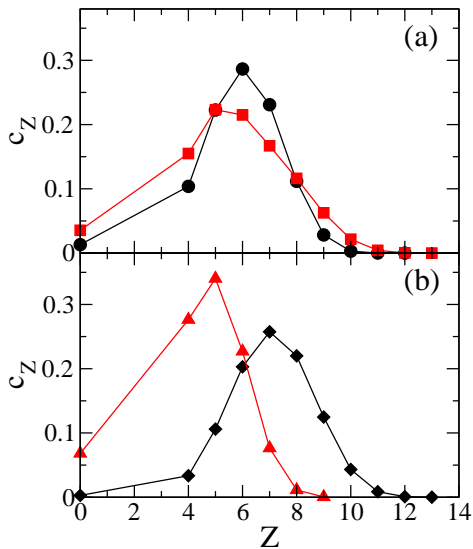}
\caption{\label{fig:fig2}
Concentration of particles with $Z$ interacting neighbors of (a) all particles measured for $R=1$ (black circles) and $R=1.4$ (red squares) at $P=10^{-4}$, and of (b) large particles (black diamonds) and small particles (red triangles) for $R=1.4$ systems shown in (a).  The lines are guide to eye.
}
\end{figure}

In marginally jammed solids, there are inevitably a fraction of particles that are
rattlers.  The concentration of rattlers $c_0$ increases with decreasing the volume
fraction approaching $\phi_c$.  For mono-disperse systems, it has been claimed that the
rattlers are necessary to keep jammed packings of hard spheres saturated \cite{donev}.
When $R$ is away from $1$, we would expect that the concentration of rattles rises,
because it is easier for small particles to stay in the vacancies formed by large network particles.  We measure $c_0(\phi)$ for each $R$ studied and extrapolate the concentration of rattlers at Point J, $c_{0c}$, as shown in Fig.~\ref{fig:fig1}(c). $c_{0c}$ increases with $R$ and finally saturates at $c_{0c}=0.5$ when $R$ is large, indicating that almost all the small particles are rattlers.  When $R\gg 1$, large particles tend to form jammed structure, while leaving small particles to stay anywhere in vacancies.  It is thus trivial to study jamming at large $R$.  From Fig.~\ref{fig:fig1}(c), we can also tell that the slope of $c_{0c}(R)$ starts to decrease with increasing $R$ approximately when $\phi_c(R)$ reaches the maximum.  Therefore, the decrease of $\phi_c$ at large $R$ indicates the trivial rattler's influence: the jammed solids are more mono-disperse like and less saturated.  In the following, we will mostly concentrate on systems with $R$ before $\phi_c(R)$ reaches the maximum.  Even though the effects of rattlers on jamming become trivial at large $R$, it is still interesting to know how the small particles affect dynamics and rheology of amorphous systems approaching jamming, since large particles are easier than small particles to jam.

The rise of $c_{0c}$ with increasing $R$ leads to a decrease of the pressure with increasing $R$,
since fewer particles are involved in interactions.  It is thus natural to imagine that the bulk modulus of marginally jammed solids, which is quantified by $P_0$, decreases with increasing $R$.
Figure~\ref{fig:fig1}(d) shows that $P_0(R)$ decreases with increasing $R$ while $L$ is kept fixed.
We express $P_0$ in units of $\epsilon/\bar{\sigma}^3$, where $\bar{\sigma}^3$
is the average particle volume, which is equivalent to fixing $L$ or particle number density $N/L^3$ when varying $R$ at fixed $\phi$ and $N$.
In the $R\rightarrow \infty$ limit, $P_0$ saturates at $P_0(1)/2$, because the system
is effectively mono-disperse with only $N/2$ network particles while the volume of the system
is kept fixed. Thus, $P_0(R)$ is solely determined by the concentration of rattlers.
As shown in Fig.~\ref{fig:fig1}(d), $P_0(R)$ is indeed proportional to $1-c_{0c}$.
We note that one can vary $R$ while keeping the diameter of the small particles $\sigma$ fixed. In
this case, $L$ increases as $R$ is increased and this increase in $L$ leads to a
decrease of $P_0$ so that $P_0$ would decrease much faster with $R$ and
eventually becomes zero in the $R \to \infty$ limit.
Special care is thus required to specify what length scale, $L$ or $\sigma$ or even $R \sigma$,
is being kept fixed when one studies how the material properties such as bulk modulus and shear
modulus vary with $R$.

\begin{figure}
\centering
\includegraphics[width=0.35\textwidth]{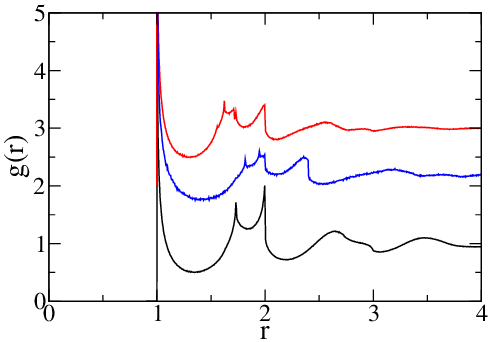}
\caption{\label{fig:fig3}
Pair correlation function $g(r)$ for jammed systems with $R=1$ (bottom, black) and $R=1.4$ at $P=10^{-4}$.  For $R=1.4$, $g(r)$ is measured for both large particles (top, red) and small particles (middle, blue).  $r$ is scaled with the particle diameter.  $g(r)$ for $R=1.4$ systems are shifted up in order to distinguish the curves.
}
\end{figure}

In Fig.~\ref{fig:fig1}(e), we show the coordination number $z$ as a function of $\phi-\phi_c$ for the same systems in Fig.~\ref{fig:fig1}(a).  Similar to what has been observed, $z=z_c+z_0(\phi-\phi_c)^b$, but $b\approx 0.6$ which is slightly different from $0.5$ as previously reported for systems with harmonic repulsion \cite{ohern}.  $z(\phi-\phi_c)$ does not show any dependence on $R$.  Therefore, the geometric properties represented by $z$ is universal for all $R$.  It has been proposed that $z$ plays a key role in determining vibrational properties of jammed solids.  The collapse of $z(\phi-\phi_c)$ over a wide range of $R$ enhances our expectation that the coordination number is a relevant order parameter to describe the jamming transition. From the scalings of $P$ and $z$ with $\phi-\phi_c$, we can easily tell that $z\propto P^{1/2}$.  Therefore, the pressure should be a better tunable parameter than the volume fraction to characterize properties of marginally jammed solids at fixed $R$.

The coordination number discussed above is a mean field value.  In marginally jammed solids, the number of interacting neighbors per particle $Z$ fluctuates over particles.  Figure~\ref{fig:fig2}(a) shows the concentration of particles with $Z$ interacting neighbors, $c_Z$ for jammed systems with $R=1$ and $1.4$.  $Z$ spans over a significant
amount of values.  For a $d-$dimensional system, a sphere needs at least $d+1$ constraints to be stable.  In Fig.~\ref{fig:fig2}(b), we separately plot $c_Z$ for small and large particles.  Apparently, large particles tend to have more constraints than small ones, because they have larger surface area to accommodate more small particles.
For the systems shown in Fig.~\ref{fig:fig2}(b), small particles have an average coordination number of $5.15$, which is much lower than the isostatic value $z_c$. Therefore, the less constrained small particles should be easier than the over constrained large particles to rearrange during plastic events when the system is under strain.  From Fig.~\ref{fig:fig2}(b) it is obvious that small particles contribute to the majority of rattlers, as discussed above.

The large deviation of $c_Z$ between large and small particles implies possible structural difference between them. Figure~\ref{fig:fig3} compares the pair correlation function $g(r)$ for the same systems in Fig.~\ref{fig:fig2}.  In mono-disperse systems, the second peak of $g(r)$ splits into two subpeaks at $r=\sqrt{3}$ and $2$ whose right hand side drops discontinuously near Point J \cite{silbert1,zou}, indicating the glass formation.  In bi-disperse systems, $g(r)$ for large particles looks similar to mono-disperse systems, except for several more subpeaks at $r<\sqrt{3}$ due to the mixing of particles with different sizes.  However, $g(r)$ for small particles is different.  The subpeaks at $r<2$ shift to the right, so it is less probable for four small particles to form the cluster corresponding to the subpeak at $r=\sqrt{3}$.

\begin{figure}
\centering
\includegraphics[width=0.4\textwidth]{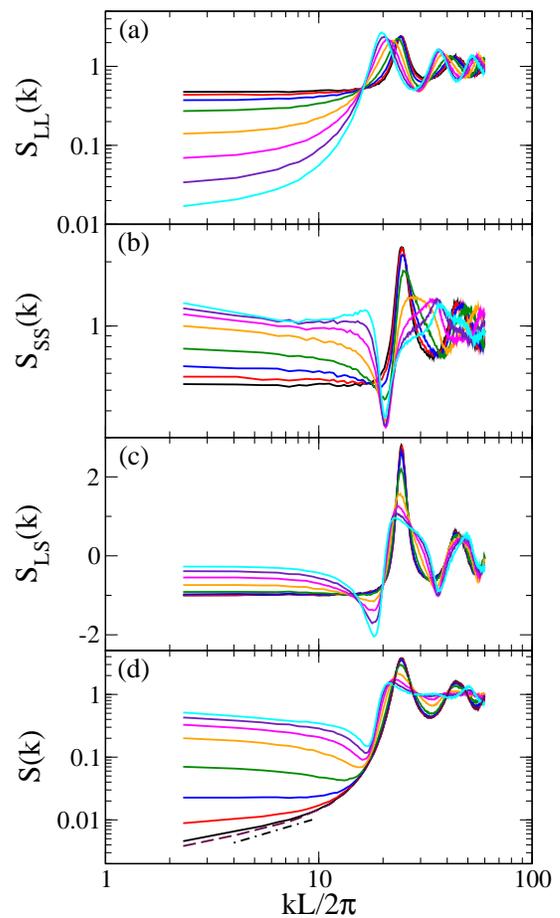}
\caption{\label{fig:fig4}
Structure factor for (a) large-large particles $S_{LL}(k)$, (b) small-small particles $S_{SS}(k)$, (c) large-small particles $S_{LS}(k)$, and (d) all particles $S(k)$ measured at $P=10^{-4}$ and at $R=1.02$ (black solid), $1.05$ (red solid), $1.1$ (blue solid), $1.2$ (green solid), $1.4$ (orange solid), $1.6$ (magenta solid), $1.8$ (purple solid), and $2.0$ (cyan solid).  The maroon dashed line in (d) is for $R=1$.  The dot-dashed line in (d) has a slope of $1$.
}
\end{figure}

The apparent variation seen in $g(r)$ reflects that the counterpart of $g(r)$ in the Fourier space, the static structure factor $S(k)$, may vary with species and $R$ as well.  In Fig.~\ref{fig:fig4}, we plot the structure factor measured for large-large particles $S_{LL}(k)$, small-small particles $S_{SS}(k)$, large-small particles $S_{LS}(k)$, and all particles $S(k)$.  Contrary to what one might have expected, $S_{LL}(k)$ and $S_{SS}(k)$ are distinct, especially in the long wave-length limit.  There are also strong dependences of $R$ on all the structure factors.  When the pressure is sufficiently low, the structure factors do not have observable pressure dependence.  Therefore, we compare the structure factors for different $R$ at the same pressure $P=10^{-4}$ without losing generality.

In the range of $R$ that we focus on ($R\in (1,2]$), $S_{LL}(k)$ is liquid like in the long wave-length limit: $S_{LL}(k)\approx S_{LL}(0)+(\zeta k)^2$, as shown in Fig.~\ref{fig:fig4}(a).  Both $S_{LL}(0)$ and $\zeta$ decreases with increasing $R$.  In thermal equilibrium systems, the structure factor at $k=0$ is proportional to the isothermal compressibility.  Although marginally jammed solids are out of equilibrium, $S_{LL}(0)$ still provides us with the information about how the large particles are spatially distributed.  When $R$ increases, large particles get closer, which leads to the decrease of the compressibility.  When $R$ is so large that almost all the small particles are rattlers, $S_{LL}(0)\approx 0$ is recovered and the jammed system is trivially equivalent to mono-disperse.

In the long wave-length limit, $S_{SS}(k)$ presented in Fig.~\ref{fig:fig4}(b) has completely different $k$ dependence from $S_{LL}(k)$.  $S_{SS}(k)$ decreases with increasing $k$ and $S_{SS}(0)$ increases with $R$.  $S_{SS}(k)$ at low $k$'s can be approximately fitted with the Ornstein-Zernike form, $S_{SS}(k)=S_{SS}(0)/\left[ 1+ \left( \xi k\right)^2\right]$, where $\xi$ increases with $R$.  When $R$ increases, viewed from small particles, the system is more dilute due to the growing amount of rattlers, which leads to an increasing compressibility and $S_{SS}(0)$.  The length $\xi$ extracted from the Ornstein-Zernike approximation of $S_{SS}(k)$ implies that small particles tend to form clusters with an average length scale $\xi$ that slightly increases with $R$.  Such a length scale would saturate to the size of the vacancies formed by large particles in the large $R$ limit.

From comparisons between Fig.~\ref{fig:fig4}(a) and (b), when $R$ is close to $1$, $S_{LL}(k)$ and
$S_{SS}(k)$ which are self-part density correlations are almost identical.  In the long
wave-length limit, they are nearly constant in $k$.  When $R\sim 1$, $S_{LS}(0)$ is negative and
cancels $S_{LL}(0)+S_{SS}(0)$ to give $S(k)\approx 0$.  As shown in Fig.~\ref{fig:fig4}(c),
$S_{LS}(0)$ increases with $R$ but remains negative in the range of $R$ studied.

The strong $R$ dependence of $S_{LL}(k)$, $S_{SS}(k)$, and $S_{LS}(k)$ indicates that what we have learned from mono-disperse jammed systems, e.g. $S(k)\sim k$, is not general and thus not special features of marginally jammed solids.  To emphasize this point, we plot the structure factor of all the particles $S(k)$ in Fig.~\ref{fig:fig4}(d).  $S(0)$ grows up with increasing $R$, which is consistent with the change of the bulk modulus with $R$ observed in Fig.~\ref{fig:fig1}.  The long wave-length behavior of $S(k)$ also strongly depends on $R$, which undergoes the transition from an increasing function to a decreasing function of $k$.  It is obvious that $S(k)\sim k$ is correct only in the $R\rightarrow 1$ limit.

\section{Conclusions}

In conclusion, by tuning the particle-size ratio $R$, we are able to study the possible effects on the properties of marginally jammed solids.  The critical volume fraction of the jamming transition is a nontrivial functions of $R$ when $R$ is not large.  Most strikingly, the density fluctuations strongly depend on $R$ and the particle species. Only jammed mono-disperse systems ($R=1$) are hyperuniform with $S(k)\sim k$.

Our observations of the material property and structural change of the jammed solids with $R$ have some interesting
implications, especially in the studies of the dynamics and rheology of amorphous systems on the approach to jamming.  Since the small and large particles have distinct density fluctuation and such a difference increases with $R$, it would be interesting to see the coexistence of jammed large particles and liquid-like small particle clusters under
thermal perturbations.  As an independent variable, $R$ may be one more parameter to control the scaling collapse of the relaxation time of glass-forming liquids \cite{xu5}.  It is also important to know whether $R$ has an effect on the equation of state, glass transition, and glass fragility of glass-forming colloidal suspensions and their response to the shear stress.

\vspace{0.1in}
\noindent {\large {\bf Acknowledgement}}
\vspace{0.1in}

We are grateful to Andrea J. Liu, Sidney R. Nagel and Lei Xu for helpful discussions and
suggestions.  This work is supported in part by Hong Kong Research Grants Council (Grant No. CUHK 400708).

\bibliographystyle{rsc}

\begin{thebibliography}{}

\bibitem{ohern} C. S. O'Hern, S. A. Langer, A. J. Liu and S. R. Nagel, {\it Phys. Rev. Lett.}, 2002, {\bf 88} 075507; C. S. O'Hern, A. J. Liu and S. R. Nagel, {\it Phys. Rev. E}, 2003, {\bf 68}, 01136.

\bibitem{liu} A. J. Liu and S. R. Nagel, {\it Nature}, 1998, {\bf 396}, 21.

\bibitem{zhang} Z. Zhang {\it et al.}, {\it Nature}, 2009, {\bf 459}, 230.

\bibitem{torquato} S. Torquato, T. M. Truskett and P. G. Debenedetti, {\it Phys. Rev. Lett.}, 2000, {\bf 84}, 2064.

\bibitem{chaud} P. Chaudhuri, L. Berthier and S. Sastry, arXiv:0910.0364, 2009.

\bibitem{xu1} N. Xu, J. Blawzdziewicz and C. S. O'Hern, {\it Phys. Rev. E}, 2005, {\bf 71}, 061306.

\bibitem{xu2} N. Xu, M. Wyart, A. J. Liu and S. R. Nagel, {\it Phys. Rev. Lett.}, 2007, {\bf 98}, 175502.

\bibitem{xu3} N. Xu, V. Vitelli, M. Wyart, A. J. Liu and S. R. Nagel, {\it Phys. Rev. Lett.}, 2009, {\bf 102}, 038001; V. Vitelli, N. Xu, M. Wyart, A. J. Liu and S. R. Nagel, submitted, arXiv:0908.2176.

\bibitem{wyart} M. Wyart, S. R. Nagel and T. A. Witten, {\it Europhys. Lett.}, 2005, {\bf 72}, 486; M. Wyart, L. E. Silbert, S. R. Nagel and T. A. Witten, {\it Phys. Rev. E}, 2005, {\bf 72}, 051306.

\bibitem{deben} P. G. Debenedetti and F. H. Stillinger, {\it Nature}, 2001, {\bf 410}, 259.

\bibitem{pohl} R. O. Pohl, X. Liu and E. Thompson, {\it Rev. Mod. Phys.}, 2002, {\bf 74}, 991.

\bibitem{donev} A. Donev, F. H. Stillinger and S. Torquato, {\it Phys. Rev. Lett.}, 2005, {\bf 95}, 090604.

\bibitem{silbert} L. E. Silbert and M. Silbert, {\it Phys. Rev. E}, 2009, {\bf 80}, 041304.

\bibitem{lbfgs} http://www.eecs.northwestern.edu/$\sim$nocedal/lbfgs.html.

\bibitem{xu4} N. Xu, submitted, arXiv:0910.0666.

\bibitem{xu5} N. Xu, T. K. Haxton, A. J. Liu and S. R. Nagel, {\it Phys. Rev. Lett.}, 2009, {\bf 103}, 245701.


\bibitem{silbert1} L. E. Silbert, A. J. Liu and S. R. Nagel, {\it Phys. Rev. E}, 2006, {\bf 73}, 041304.

\bibitem{zou} L.-N. Zhou {\it et al.}, {\it Science}, 2009, {\bf 326}, 408.

\end{thebibliography}

\end{document}